\documentclass[a4paper]{article}
\usepackage{dingbat}
\usepackage{Odyssey2020}
\usepackage{epsfig,amssymb,amsmath}
\usepackage{url}
\usepackage{hyperref}
\usepackage{bm}
\usepackage{upgreek}
\usepackage{subcaption,siunitx,booktabs}
\usepackage[table]{xcolor}
\usepackage{subcaption}

\ninept

\newcommand{\ie}{\textit{i}.\textit{e}., }
\newcommand{\eg}{\textit{e}.\textit{g}. }

\newcommand{\N}{\mathcal{N}}
\newcommand{\T}{^\mathsf{T}}

\renewcommand{\vec}[1]{\bm{#1}}
\newcommand{\mat}[1]{\mathbf{#1}}

\newcommand\nextline{\raisebox{0.8ex}[0pt][0pt]{\reflectbox{\carriagereturn}}}

\let\OLDthebibliography\thebibliography
\renewcommand\thebibliography[1]{
  \OLDthebibliography{#1}
  \setlength{\parskip}{0pt plus 0.2ex}
}

\newcommand\blfootnote[1]{%
  \begingroup
  \renewcommand\thefootnote{}\footnote{#1}%
  \addtocounter{footnote}{-1}%
  \endgroup
}

\name{Ville Vestman$^{1}$, Kong Aik Lee$^2$, Tomi H. Kinnunen$^1$}

\title{Neural i-vectors}

\address{
  $^1$Computational Speech Group, University of Eastern Finland, Finland\\
  $^2$Biometrics Research Laboratories, NEC Corporation, Japan \\
  {\small \tt{vvestman@cs.uef.fi, kongaik.lee@nec.com, tkinnu@cs.uef.fi}}}

\begin{document}

\setlength{\floatsep}{10pt plus 6.0pt minus 2.0pt}
\setlength{\dblfloatsep}{6pt plus 2.0pt minus 2.0pt}
\setlength{\textfloatsep}{6pt plus 2.0pt minus 2.0pt}
\setlength{\dbltextfloatsep}{10pt plus 2.0pt minus 2.0pt}
\setlength{\intextsep}{8pt plus 2.0pt minus 2.0pt}

\setlength{\abovedisplayskip}{3pt plus 3.0pt minus 1.0pt}
\setlength{\belowdisplayskip}{3pt plus 3.0pt minus 1.0pt}
\setlength{\abovedisplayshortskip}{-2pt plus 1.0pt minus 1.0pt}
\setlength{\belowdisplayshortskip}{2pt plus 1.0pt minus 1.0pt}

\maketitle

\begin{abstract}

Deep speaker embeddings have been demonstrated to outperform their generative counterparts, i-vectors, in recent speaker verification evaluations. To combine the benefits of high performance and generative interpretation, we investigate the use of deep embedding extractor and i-vector extractor in succession. 
To bundle the deep embedding extractor with an i-vector extractor, we adopt aggregation layers inspired by the Gaussian mixture model (GMM) to the embedding extractor networks. The inclusion of GMM-like layer allows the discriminatively trained network to be used as a provider of sufficient statistics for the i-vector extractor to extract what we call \emph{neural i-vectors}.
We compare the deep embeddings to the proposed neural i-vectors on the Speakers in the Wild (SITW) and the Speaker Recognition Evaluation (SRE) 2018 and 2019 datasets. On the core-core condition of SITW, our deep embeddings obtain performance comparative to the state-of-the-art. The neural i-vectors obtain about 50\% worse performance than the deep embeddings, but on the other hand outperform the previous i-vector approaches reported in the literature by a clear margin.

\end{abstract}

\section{Introduction}
\emph{Automatic speaker verification} (ASV) systems extract speaker-related information from a pair of speech recordings (enrollment and test) to decide whether the speakers in the two recordings are the same. This is done by computing similarity score between speaker-related features in the two recordings. While the base features have remained the same for decades \cite{Davis80}, extraction and comparison of speaker traits from these features has coevolved with advances in machine learning. Much of ASV research has focused on modeling low-level speech feature distributions via \emph{Gaussian mixture models} (GMMs) \cite{Reynolds95_speaker,Reynolds2000_gmmubm,Kenny2007-jfa-vs-eigenchannel,Dehak10frontend}. Common to models such as GMM with \emph{universal background model} (GMM-UBM) \cite{Reynolds2000_gmmubm}, \emph{joint factor analysis} (JFA) \cite{Kenny2007-jfa-vs-eigenchannel} and \emph{i-vector} \cite{Dehak10frontend} is the use of GMM to model acoustic features within recording(s). 

What has changed throughout the years, however, is how  speaker comparison is carried out. In the classic GMM pipelines \cite{Reynolds95_speaker,Reynolds2000_gmmubm}, features in the enrollment utterance(s) are used to train a speaker-dependent GMM, and comparison consists of evaluating the likelihoods of the target speaker model and the UBM to form an average log-likelihood ratio over all frames. In contrast to these \emph{frame-based} approaches, the modern approach is to first represent the enrollment and test utterances as vectors of the same dimensionality. They can then be compared using a simple inner product, or a trainable classifier~\cite{matvejka2011full}. How these vectors are defined (and called) has changed throughout the years. The early approaches, driven by the success of GMMs, used high-dimensional GMM \emph{supervectors} \cite{campbell2006support} with inner product scoring, typically implemented using \emph{support vector machines} (SVMs). Through base work in \cite{Kenny2007-jfa-vs-eigenchannel}, this was followed up by the highly-successful \emph{i-vector} framework \cite{Dehak10frontend} where GMM supervectors are presented as points in a low-dimensional latent subspace. Following trends in deep learning, the focus has recently shifted towards deep neural network (DNN) based features \cite{snyder2018x}, called nowadays \emph{embeddings}. The idea to represent utterances as vectors, however, is the same as before, with the same back-end classifiers  \cite{PrinceElder2007} used with GMM- and neural network based embeddings. 


As the title suggests, we focus on i-vector extraction along the lines of classic GMM-based pipelines, but with a `neural twist'. The general idea, of course, is not new. The three building blocks of any GMM-based method are (a) \textbf{a frame-level feature extractor} (\eg MFCC extractor), (b) \textbf {a dictionary} (\eg a UBM), and (c) \textbf{a posterior estimator} (\eg feature vector alignment to dictionary components), each of which has been successfully replaced in prior work by their neural versions \cite{tian2015investigation, lei2014novel}. In contrast to these studies that have focused either on replacing one or two of the components only, or using GMM-inspired components \cite{cai2018novel, xie2019utterance, chen2019tied} to implement neural embedding extractors, we obtain all the three as `side-products' from a neural network and proceed with conventional i-vector extractor training on top of them. Noting that (a), (b) and (c) are the only needed building blocks of \emph{any} GMM-based embedding --- be it a GMM-UBM \cite{Reynolds2000_gmmubm}, JFA \cite{Kenny2007-jfa-vs-eigenchannel}, GMM-supervector \cite{campbell2006support}, or i-vector \cite{Dehak10frontend} --- this opens up a pathway to re-address any of the classic pipelines, still respecting the undeniable performance gains demonstrated by the recent neural approaches. 

Our focus on i-vectors is arbitrary and the goal of our work is \emph{not} to improve upon state-of-the-art in deep neural network based speaker embeddings. Instead, we aim to 
demonstrate that classic GMM-based ASV pipelines may not be inferior because of their model structures \emph{per se}, but in the adoption of generic (nondiscriminative) elements. Classic frame-based GMM approaches have certain, nearly forgotten advantages, such as the ability to provide `partial' scores at a fine temporal scale --- the frame level. This might be particularly useful in speaker diarization (not addressed here) and speaker recognition from short utterances. 
Even if DNN embeddings appear to perform well in short duration ASV tasks \cite{snyder2019speaker}, we argue that using GMMs retains all the benefits of generative modeling, such as the possibility to do sampling and obtaining uncertainty estimates for features and speaker embeddings. These add up to transparency and explainability demanded with increasing frequency from any machine learning system.

\section{Modern speaker embedding extractors}
\begin{figure*}[t]
  \centerline{\includegraphics[trim={2.0cm 11.3cm 2.0cm 1.4cm}, clip, width=0.8\linewidth]{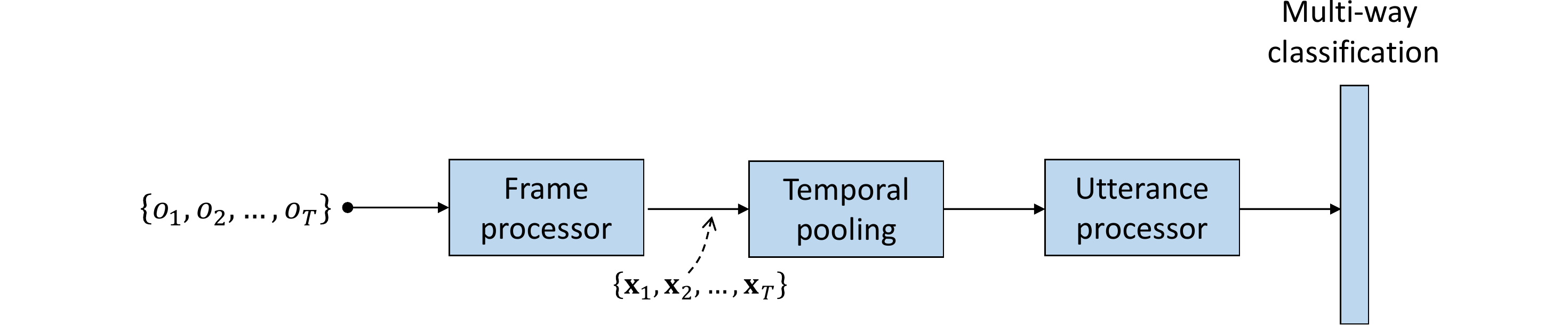}}
  \caption{{\it An x-vector extractor consists of three functional blocks: a frame-level processor, a temporal pooling layer, and classifier. X-vector embeddings are derived from the affine transformation after the pooling layer.}}
  \label{fig:x-vector}
\end{figure*}

Deep neural networks used for extracting speaker discriminative embeddings typically consist of three main parts (see Figure \ref{fig:x-vector}). The first part of the network operates on \emph{frame-level features} as an input in order to construct discriminative features from short time contexts, ranging from $100$ milliseconds up to a few seconds. The frame-level layers are followed by the second main component, \emph{temporal aggregation layer}, which converts the variable length input feature vector sequence to a fixed-dimensional representation. Finally, the last part of the network, which consists of one or two feedforward layers and the output layer, acts as a \emph{classifier} for speaker identities. The speaker embeddings are usually extracted from the first fully connected layer after the aggregation layer \cite{snyder2018x}.

Each of the three main parts can be implemented in multiple different ways. The frame-level component is often implemented as 1D convolutional neural network (CNN) \cite{zeinali2019improve}, 2D CNN \cite{chung2018voxceleb2}, or as some variant of time-delay neural network \cite{villalba2020}. In 1D CNN, the convolution kernel slides over the temporal dimension (frames), whereas in 2D convolution, the kernel slides over both time and frequency dimensions.

\begin{figure*}[t]
  \centering
  \centerline{\includegraphics[trim={2.0cm 10.9cm 2.0cm 1.4cm}, clip, width=0.8\linewidth]{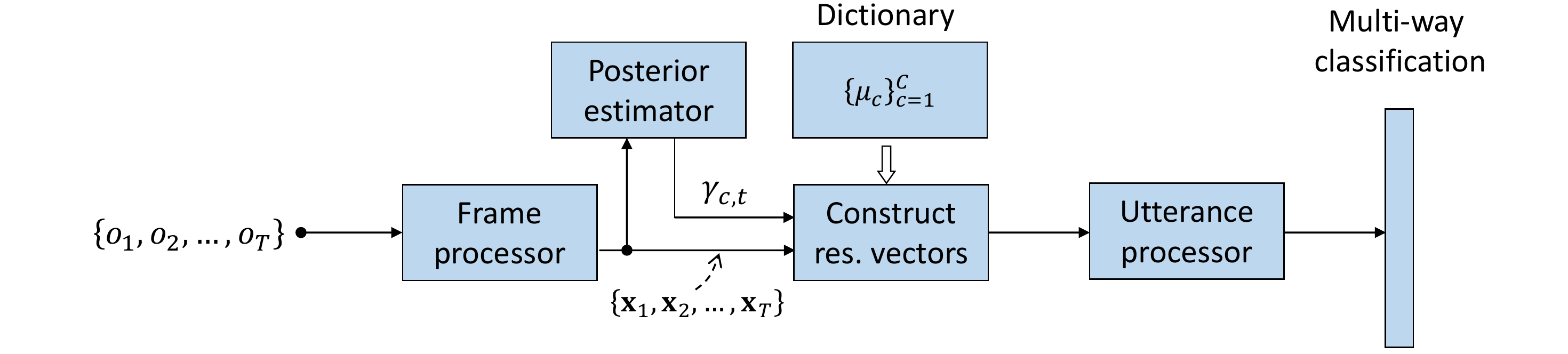}}
  \caption{{\it The centrepiece of learnable dictionary encoder (LDE) and NetVLAD is the frame processor, frame posterior estimator and dictionary that are trained jointly to minimize a classification loss.}}
  \label{fig:cluster-pool}
\end{figure*}

There are two commonly used approaches for temporal aggregation. In the first approach \cite{snyder2018x}, relatively \emph{high-dimensional} features are obtained from CNN/TDNN, which are then aggregated by computing the (sample) mean and the standard deviation of the feature vectors over time. The output of the aggregation layer is then formed by concatenating the mean and standard deviation vectors. The second approach \cite{cai2018novel} of aggregation assumes relatively \emph{low-dimensional} features (akin to conventional hand-crafted acoustic features) from CNN/TDNN but assigns them into multiple clusters (see Figure \ref{fig:cluster-pool}). Here, the aggregation is performed for each cluster separately, resulting in locally aggregated descriptor vectors. Finally, the locally aggregated descriptors are concatenated to form a higher-dimensional residual vector. This approach is analogous to the process how GMM mean supervectors are formed in the GMM-UBM framework.

\begin{figure*}[h!]
  \centering
  \centerline{\includegraphics[trim={2.0cm 10.9cm 2.0cm 1.9cm}, clip, width=0.8\linewidth]{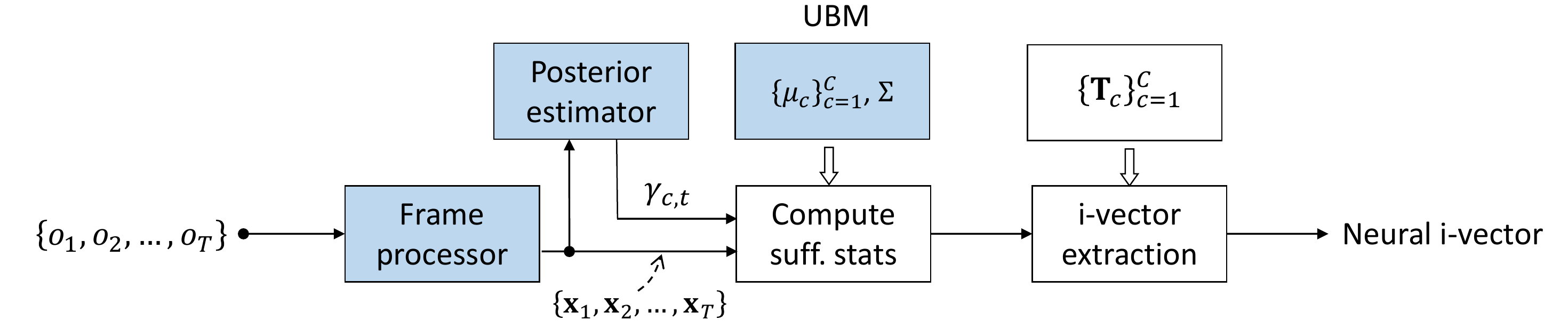}}
  \caption{{\it The proposed neural i-vector relies on a deep structured front-end (shaded boxes) to extract sufficient statistics, which are then used for generative embedding.}}
  \label{fig:neural-ivector}
\end{figure*}

\begin{figure*}[h!]
  \centering
  \centerline{\includegraphics[trim={2cm 12.3cm 2cm 2.0cm}, clip, width=0.8\linewidth]{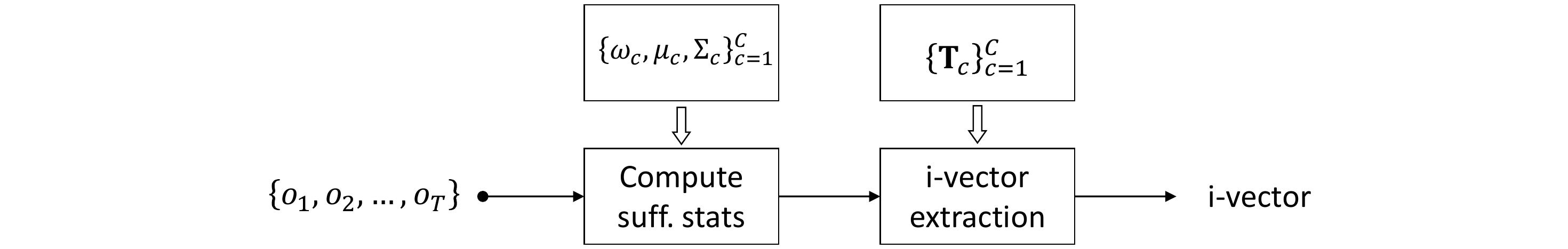}}
  \caption{{\it An i-vector extractor is built upon a Universal Background Model (UBM) defined by the parameter set consists of weights, mean vectors, and covariance matrices.}}
  \label{fig:ivector}
\end{figure*}

\vspace{-2mm}
\section{Cluster-wise temporal aggregation}
\vspace{-1mm}
We focus on cluster-wise temporal aggregation methods as they offer a natural pathway to utilize GMM-based speaker verification approaches, such as the i-vector approach, together with discriminatively trained features. In the following, we consider two recent aggregation methods known as \emph{learnable dictionary encoder} (LDE) \cite{zhang2017deep, cai2018novel} and \emph{NetVLAD} \cite{arandjelovic2016netvlad, xie2019utterance}, where VLAD is an acronym for ``\underline{v}ector of \underline{l}ocally \underline{a}ggregated \underline{d}escriptors''. As we will show below, both can be regarded as discriminatively trained GMM-supervector \cite{campbell2006support} encoders with specific assumptions.

Let us first recall the formula for \emph{posterior} computation of a Gaussian mixture component given a feature vector $\vec{x}_t$ (time index $t=1,\ldots,T$). By letting $\vec{\theta} = \{\vec{\mu}_c, \mat{\Sigma}_c, w_c\}_{c=1}^C$ be a GMM of $C$ components with mean vectors $\vec{\mu}_c$, covariance matrices $\mat{\Sigma}_c$, and component weights $w_c$, we can compute the posteriors as follows:
\begin{equation}
\label{eq:gmm_posterior}
    \gamma_{c,t} = P(c|\vec{x}_t) = \frac{w_c\N(\vec{x}_t|\vec{\mu}_c, \mat{\Sigma}_c)}{\sum_{l=1}^{C} w_l\N(\vec{x}_t|\vec{\mu}_l, \mat{\Sigma}_l)}, \quad c=1,\ldots,C.
\end{equation}
By denoting
\begin{equation}
    \beta_c = \log \left( \frac{w_c}{\sqrt{(2\pi)^D|\mat{\Sigma}_c|}} \right),
\end{equation}
where $D$ is the dimension of feature vectors, we can expand \eqref{eq:gmm_posterior} to form
\begin{equation}
    \label{eq:gmm_posterior_expanded}
    \gamma_{c,t} = \frac{\exp \left[-\frac{1}{2} (\vec{x}_t - \vec{\mu}_c)\T \mat{\Sigma}_c^{-1} (\vec{x}_t - \vec{\mu}_c) + \beta_c \right]}{\sum_{l=1}^{C} \exp \left[-\frac{1}{2} (\vec{x}_t - \vec{\mu}_l)\T \mat{\Sigma}_l^{-1} (\vec{x}_t - \vec{\mu}_l) + \beta_l \right]}.
\end{equation}

\subsection{Learnable dictionary encoder}
Equation \eqref{eq:gmm_posterior_expanded} holds for any GMM with unrestricted covariance matrices. In the following, we consider special cases, where the covariance matrices are restricted to have specific forms.
First, by assuming \emph{isotropic} covariance matrices (\ie $\mat{\Sigma}_c = s_c \mat{I}$, with $s_c > 0$), \eqref{eq:gmm_posterior_expanded} becomes
\begin{equation}
    \label{eq:lde_posteriors}
    \gamma_{c,t} = \frac{\exp \left[ -\frac{1}{2} s_c \Vert \vec{x}_t - \vec{\mu}_c \Vert^2 + \beta_c \right]}{\sum_{l=1}^{C} \exp \left[ -\frac{1}{2} s_l \Vert \vec{x}_t - \vec{\mu}_l \Vert^2 + \beta_l \right]},
\end{equation}
where $\Vert \cdot \Vert$ denotes the Euclidean norm. This is the formulation used for posterior computation in \cite{villalba2020} with LDE. Earlier works on LDE \cite{zhang2017deep, cai2018novel}, did not include the bias terms $\beta_c$. Both the parameters $s_c$ that define the isotropic covariance matrices, and the bias terms $\beta_c$ as well as the cluster centroids $\vec{\mu}_c$ are learnable parameters of the LDE layer.

After computing the posteriors, the construction of the output of LDE layer is a two step process. First, the input features are temporally aggregated with respect to each cluster. This is done by computing the weighted means $\vec{m}_c$ of residuals $\vec{\mu}_c - \vec{x}_t$ around the cluster centroids for each cluster $c$:
\begin{equation}
    \label{eq:lde_clusterwise_representation}
    \vec{m}_c = \frac{\sum_{t=1}^T \gamma_{c,t} (\vec{\mu}_c - \vec{x}_t)}{\sum_{t=1}^T \gamma_{c,t}}.
\end{equation}
The second step is to concatenate the cluster-wise representations to form a supervector \mbox{$\vec{m} = (\vec{m}_1\T, \vec{m}_2\T, \ldots \vec{m}_T\T)\T$}, which is the output of the LDE layer.

While the original formulation of LDE uses separate \emph{isotropic} covariance matrices for each component, it is straightforward to modify the LDE layer to operate with \emph{diagonal} covariance matrices, or to use one \emph{shared diagonal} or \emph{spherical} covariance matrix for all components. In our experiments, we consider only the shared diagonal matrix formulation besides the original formulation with non-shared spherical covariances to limit the computational burden.


\renewcommand*{\arraystretch}{1.0}
\begin{table}[t]
\caption{{\it Comparison of LDE and NetVLAD.} \label{table:lde_vs_netvlad}}
\vspace{-1mm}
\centering
\begin{tabular}{l c c}
\toprule
Computation step & LDE & NetVLAD \\
\midrule
Posterior computation & Eq. \eqref{eq:lde_posteriors} & Eq. \eqref{eq:netvlad_posteriors} \\
Cluster-wise representations & Eq. \eqref{eq:lde_clusterwise_representation} & Eq. \eqref{eq:netvlad_clusterwise_representation}\\
Supervector normalization & --- & Length-norm\\
\bottomrule
\centering
\vspace{-3mm}
\end{tabular}
\end{table}
\renewcommand*{\arraystretch}{1.0}

\subsection{NetVLAD encoder}
Let us next assume \emph{shared full covariance matrices} (\ie $\mat{\Sigma}_c = \mat{\Sigma}$ $\forall c$), which will lead to the NetVLAD formulation of posterior computation. The shared covariance assumption simplifies \eqref{eq:gmm_posterior_expanded} to
\begin{equation}
     \gamma_{c,t} = \frac{\exp \left[\vec{\mu}_c\T \mat{\Sigma}^{-1} \vec{x}_t + \log(w_c) - \frac{1}{2} \vec{\mu}_c\T \mat{\Sigma}^{-1} \vec{\mu}_c\right]}{\sum_{l=1}^{C} \exp \left[\vec{\mu}_l\T \mat{\Sigma}^{-1} \vec{x}_t + \log(w_l) - \frac{1}{2} \vec{\mu}_l\T \mat{\Sigma}^{-1} \vec{\mu}_l\right]},
\end{equation}
which allows us to write
\begin{equation}
\label{eq:netvlad_posteriors}
    \gamma_{c,t} = \frac{\exp \left[\vec{\omega}_c\T \vec{x}_t + \psi_c \right]}{\sum_{l=1}^{C} \exp \left[\vec{\omega}_l\T \vec{x}_t + \psi_l \right]},
\end{equation}
where
\begin{equation*}
    \vec{\omega}_c = \mat{\Sigma}^{-1} \vec{\mu}_c \quad \textrm{and} \quad \psi_c = \log(w_c) - \frac{1}{2} \vec{\mu}_c\T \mat{\Sigma}^{-1} \vec{\mu}_c.
\end{equation*}
Equation \eqref{eq:netvlad_posteriors} can be implemented as an affine transform followed by softmax operation over clusters, which is exactly what is done in NetVLAD layer to compute the posteriors. The NetVLAD layer has $\vec{\omega}_c$, $\psi_c$, and $\vec{\mu}_c$ as its learnable parameters. In terms of the number of learnable parameters, the correspondence between NetVLAD and GMM  with shared covariances is not exact as covariance matrix $\mat{\Sigma}$ contains $D(D+1)/2$ free parameters, whereas a matrix containing all $\vec{\omega}_c$ vectors has $CD$ parameters.

In NetVLAD, the construction of the output supervector differs from LDE in two ways. First, NetVLAD length-normalizes the component-wise outputs:
\begin{equation}
    \label{eq:netvlad_clusterwise_representation}
    \vec{m}_c = \frac{\sum_{t=1}^T \gamma_{c,t} (\vec{\mu}_c - \vec{x}_t)}{\left \Vert \sum_{t=1}^T \gamma_{c,t} (\vec{\mu}_c - \vec{x}_t) \right \Vert}.
\end{equation}
The second difference is that the supervector $\vec{m}$, obtained by concatenating the cluster-wise outputs $\vec{m}_c$, is further length-normalized to unit sphere. 

The differences between LDE and NetVLAD are summarized in Table \ref{table:lde_vs_netvlad}. As LDE and NetVLAD differ in the posterior computation as well as whether or not length-normalizations are applied, it is challenging to identify the potential causes of the performance difference between the two methods (if such difference is to exist). Therefore, we also study a hybrid approach (referred as NetVLAD/LDE), in which the posterior computation follows the NetVLAD approach, while the rest of the steps follow the LDE approach.

\section{Utilizing aggregation statistics for i-vector extraction}

Deep speaker embedding \cite{snyder2018x} has been demonstrated to outperform the i-vector representation shown in Figure \ref{fig:ivector}. The enhanced performance is attained by (1) training the network using large amount of training data via \emph{data augmentation}, and (2) \emph{discriminative training} (e.g., multi-class cross entropy cost, angular margin cost \cite{liu2017sphereface}). The drawback is lack of generative interpretation. We propose to combine the benefits from both sides, leading to the so-called \textbf{neural i-vector} shown in Figure~\ref{fig:neural-ivector}. In the neural i-vector, we utilize the features, posterior estimator, and the UBM that all have been trained discriminatively using speaker labels. This differs from the DNN i-vector presented in \cite{lei2014novel} as it requires senone labels and does not utilize discriminatively trained features.

To extract neural i-vectors, we do not compute \eqref{eq:lde_clusterwise_representation} or \eqref{eq:netvlad_clusterwise_representation}, but instead compute the suffiecient statistics in a standard way \cite{kenny2012small} as follows: 
\begin{align}
    z_c &= \sum_{t=1}^T \gamma_{c,t}, \label{eq:zeroth_stats} \\
    \vec{f}_c &= \sum_{t=1}^T \gamma_{c,t} \vec{x}_t, \label{eq:first_stats} \\
    \mat{S}_c &= \sum_{t=1}^T \gamma_{c,t} \vec{x}_t \vec{x}_t\T. \label{eq:second_stats}
\end{align}
Here the features and posteriors are extracted from the embedding extractor network. The obtained statistics can be then easily used with any available i-vector code to train the i-vector extractor and to extract the i-vectors.

\vspace{-1mm}
\section{Speaker verification experiments}

\subsection{Network architectures and training procedure}

The network architectures designed for this study are all derived from the standard x-vector architecture presented in \cite{snyder2018x}. Our most elementary architecture differs from \cite{snyder2018x} in the following ways:
\begin{itemize}
    \item As in \cite{zeinali2019improve}, we use \emph{non-dilated} 1D CNN instead of TDNN with dilations used in \cite{snyder2018x}.
    \item As in \cite{zeinali2019improve}, we use \emph{leaky} rectified linear unit (LReLU) activations (with slope of $0.01$) instead of ReLUs.
    \item We have only one embedding layer (rather than two) after the aggregation layer. In our preliminary experiments, we did not find adding another layer to decrease the resulting speaker verification equal error rates.
\end{itemize}

\begin{figure*}[h!]
  \centering
  \centerline{\includegraphics[trim={2.0cm 12.2cm 2.0cm 1.9cm}, clip, width=0.8\linewidth]{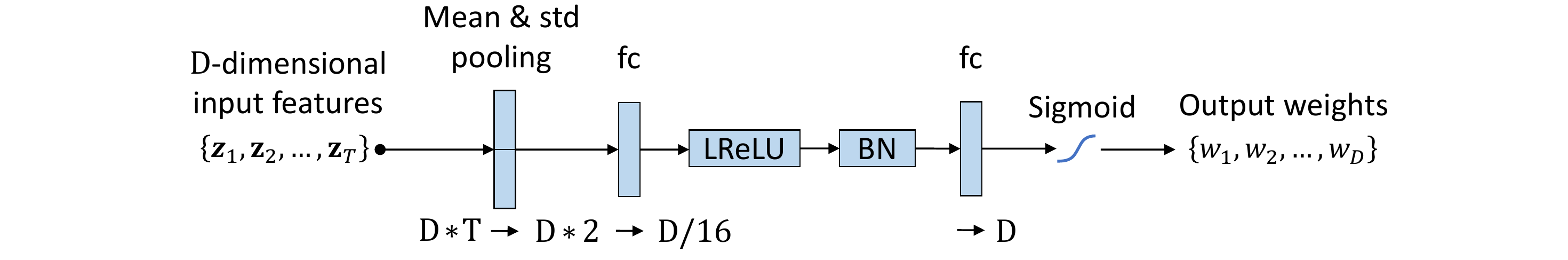}}
  \caption{{\it The squeeze-and-excitation (SE) module. The output weights are used to weight the input features as shown in Figs. \ref{fig:tdnn-se} and~\ref{fig:tdnn-res-se}.}}
  \label{fig:se}
\end{figure*}

\begin{figure*}[h!]
  \centering
  \centerline{\includegraphics[trim={2.0cm 12.7cm 2.0cm 1.9cm}, clip, width=0.8\linewidth]{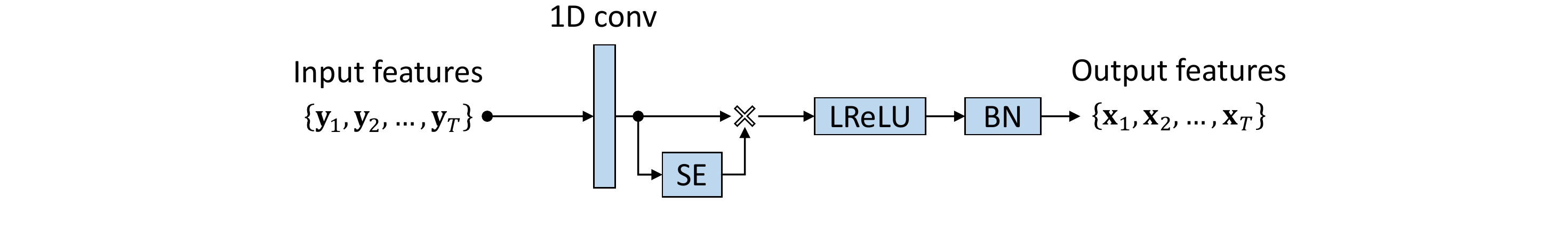}}
  \caption{{\it A TDNN module with squeeze-and-excitation (SE). This module is used to build the frame processor of TDNN-SE network.}}
  \label{fig:tdnn-se}
\end{figure*}

\begin{figure*}[h!]
  \centering
  \centerline{\includegraphics[trim={2.0cm 11.6cm 2.0cm 3.2cm}, clip, width=0.8\linewidth]{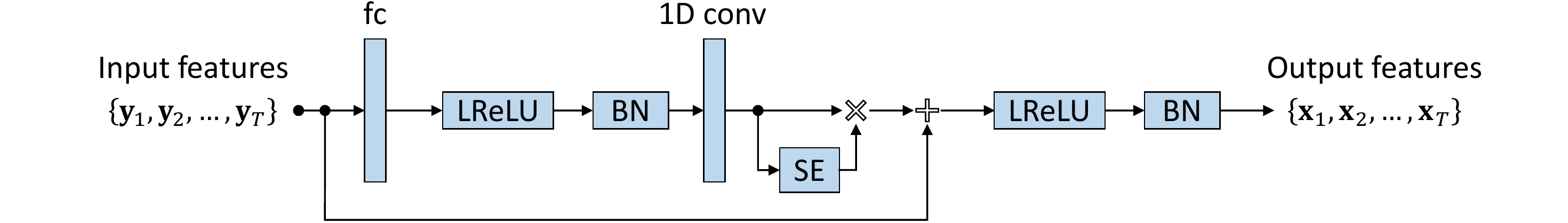}}
  \caption{{\it A Residual module with squeeze-and-excitation (SE). This module is used to build the frame processor of TDNN-RES-SE network. Compared to the module in Figure \ref{fig:tdnn-se}, this module adds a fully connected layer (fc) and a residual connection. The residual connection adds input features to the features obtained from the SE operation.}}
  \label{fig:tdnn-res-se}
\end{figure*}

We extend our default network (referred as TDNN) by adding \emph{squeeze-and-excitation} (SE) modules \cite{hu2018squeeze} to the TDNN layers. The SE module aims to improve the representative power of hidden features by reweighting them using information from global temporal statistics of features. Using the terminology of \cite{roy2018concurrent}, we adopt SE modules to perform \emph{temporal squeeze} and \emph{channel (feature) excitation}. That is, the output features of 1D CNN layer are weighted by factors computed from temporally pooled (non-weighted) features. Our implementation of the SE module is depicted in Figure \ref{fig:se}, while Figure \ref{fig:tdnn-se} illustrates how the SE module is added to the TDNN layer. The resulting TDNN-SE network architecture is presented in Table \ref{table:se_architecture}. Our SE module differs from the original in that it computes standard deviations in addition to means during the squeeze phase. In the excitation phase, we add batch normalization between the fully connected layers, as shown in Figure~\ref{fig:se}.

Inspired by the widely used ResNet architecture \cite{he2016deep}, our next network variant includes \emph{residual} modules. Our implementation of a residual module (referred as TDNN-RES-SE) includes a fully connected layer, a 1D convolutional layer, and a SE module, as depicted in Figure \ref{fig:tdnn-res-se}. The network architecture is shown in Table \ref{table:resse_architecture}. We replace neither the first nor the last TDNN-SE layer with the residual modules, as residual modules require the number of input and output features to be the same. The first layer has relatively low-dimensional MFCCs as its input, while the output size of the last layer depends on the aggregation method used. Networks with the mean and standard deviation pooling produce $1500$-dimensional feature vectors at the output of the last TDNN layer. Networks with LDE or NetVLAD, in turn, produce TDNN outputs of $128$-dimensions. With LDE and NetVLAD, we use $64$ clusters, resulting in $8192$-dimensional output vectors from the aggregation layer.

\renewcommand*{\arraystretch}{1.0}
\begin{table}[t]
\caption{{\it The architecture of TDNN-SE network.} \label{table:se_architecture}}
\vspace{-2mm}
\centering
\begin{tabular}{l l c c}
\toprule
\# & Layer type & CNN kernel size & Output dim. \\
\midrule
1 & TDNN-SE & 5 & 512 \\
2 & TDNN-SE & 5 & 512 \\
3 & TDNN-SE & 7 & 512 \\
4 & TDNN-SE & 1 & 512 \\
5 & TDNN-SE & 1 & 1500 \\
6 & Aggregation & --- & 3000 \\
7 & FC-LReLU-BN & --- & 512 \\
8 & FC-softmax & --- & \#speakers \\
\bottomrule
\centering
\vspace{-3mm}
\end{tabular}
\end{table}
\renewcommand*{\arraystretch}{1.0}

\renewcommand*{\arraystretch}{1.0}
\begin{table}[t]
\caption{{\it The architecture of TDNN-RES-SE network. The output sizes of the last TDNN-SE layer and the aggregation layer depend on the aggregation method. If aggregation using means and standard deviations is used, these sizes are $1500$ and $3000$, but if LDE or NetVLAD is used, the sizes are $128$ and $8192$.} \label{table:resse_architecture}}
\vspace{-2mm}
\centering
\begin{tabular}{l l c c}
\toprule
\# & Layer type & CNN kernel size & Output dim. \\
\midrule
1 & TDNN-SE & 5 & 512 \\
2,3 & TDNN-RES-SE & 5 & 512 \\
4,5 & TDNN-RES-SE & 7 & 512 \\
6,7 & TDNN-RES-SE & 1 & 512 \\
8 & TDNN-SE & 1 & 1500/128 \\
9 & Aggregation & --- & 3000/8192 \\
10 & FC-LReLU-BN & --- & 512 \\
11 & FC-softmax & --- & \#speakers \\
\bottomrule
\centering
\vspace{-3mm}
\end{tabular}
\end{table}
\renewcommand*{\arraystretch}{1.0}

All our networks are implemented with PyTorch \cite{paszke2019pytorch}. The Kaldi toolkit \cite{povey2011kaldi} is used to extract speech activity labels and $60$-dimensional MFCCs (without delta features), used as the input features. PyKaldi \cite{can2018pykaldi} is used to load the features in Kaldi format in Python and to perform \emph{cepstral mean normalization} (CMS) for the features.

For network training, we use four second long segments selected from random positions of the training utterances. During training, we feed about \num[group-separator={\text{\,}}]{14000} short segments from each training speaker to the network in minibatches of size $64$. Network weights are updated to minimize cross-entropy loss using stochastic gradient descend optimizer with weight decay parameter set to $0.001$. We use a learning rate schedule that decreases the learning rate from $0.05$ to $0.0002$ during the training.

\vspace{-1mm}
\subsection{Neural i-vector training details}
\vspace{-1mm}
We utilized the \emph{augmented} form of i-vector extractor as described in \cite{vestman2019}\footnote{The PyTorch re-implementation of Kaldi's i-vector extractor used in our study is available at \url{https://github.com/vvestman/pytorch-ivectors}}. In the augmented form, the UBM mean vectors are augmented into the first column of the \emph{total variability matrix}~$\mat{T}$ and they are thus updated after the each iteration of extractor training, unlike in the standard formulation. For modeling residual covariances in the total variability model, we used a diagonal covariance matrix that was shared between all components. To initialize the first column of~$\mat{T}$ and the residual covariance matrix, we used means and covariances computed from the sufficient statistics \eqref{eq:zeroth_stats}, \eqref{eq:first_stats}, and \eqref{eq:second_stats} of the training data. We set the i-vector dimension to $512$, which is the same as the dimension of the network embeddings.

\subsection{Training data}
\vspace{-1mm}
To train the neural networks, i-vector extractors, and scoring back-ends, we used 16 kHz speech data from VoxCeleb1 \cite{nagrani2017voxceleb} and VoxCeleb2 \cite{chung2018voxceleb2}. VoxCeleb data has been collected from YouTube by automatic means. Like in \cite{villalba2020}, we concatenated all the segments that were extracted from the same YouTube source video, and used these concatenated segments as the training data. We excluded all concatenated segments less than six seconds long. After filtering out the short segments, we were left with data extracted from \num[group-separator={\text{\,}}]{149754} unique YouTube videos, containing $7365$ speakers. This data was then augmented five-fold using Kaldi's augmentation recipe, resulting in total of \num[group-separator={\text{\,}}]{748770} concatenated segments. The augmentation creates copies of data by reverberating speech or by adding noise, babble, or music to the speech.

\vspace{-1mm}
\subsection{Evaluation data and metrics}
\vspace{-1mm}
We evaluated all the ASV systems on \emph{Speakers in the Wild} (SITW) \cite{mclaren2016speakers} and \emph{NIST Speaker Recognition Evaluation} (SRE) 2018 \cite{sre18} and 2019 \cite{sre19} data. From SITW, we selected core-core (referred here as `core') and core-multi (referred as `multi') conditions. In both, only a single speaker appears in each of the enrollment segments, but in the multi condition, the test utterances may contain speech from multiple speakers (unlike in the core condition). The core condition evaluation contains \num[group-separator={\text{\,}}]{721788} trials, out of which $3658$ are target trials. The multi condition contains \num[group-separator={\text{\,}}]{2010683} trials, out of which \num[group-separator={\text{\,}}]{10045} are target trials.

The SRE 2018 and the SRE 2019 both consist of two separate evaluations. One is based on telephone speech data in Call My Net 2 (CMN2) corpus, while the other one is based on Video Annotation for Speech Technology (VAST) corpus. In this study, we evaluated only the VAST portions of SREs as VAST data is a better match to our VoxCeleb training data. The SRE 2018 evaluation contains \num[group-separator={\text{\,}}]{31815} trials, out of which $315$ are target trials, while the SRE 2019 has \num[group-separator={\text{\,}}]{67348} trials, out of which $452$ are target trials.

With SREs, we used the diarization labels provided by NIST for the enrollment side to remove the unwanted portions of speech from the enrollment. We did not perform diarization of the test side for any of the datasets.

For each set of evaluation trials, we report \emph{equal error rate} (EER) and normalized minimum detection cost (minDCF). See~\cite{Sadjadi2019} for details of minDCF. We adopted the same minDCF parameters as used in SRE 2018 and 2019 evaluations. That is, we set the costs of miss and false alarm equal to one ($C_{\textrm{miss}}$ = $C_{\textrm{fa}} = 1$), and the target prior $P_\textrm{target}$ to $0.05$.

\addtolength{\tabcolsep}{-1pt}  
\renewcommand*{\arraystretch}{1.1}
\begin{table*}[t]
\caption{\label{table:results} {\it Speaker verification results for the systems evaluated in this study. In addition to the deep speaker embedding systems, the results are reported for four neural i-vector systems each of which are based on different variations of the aggregation layer.}}
\centerline{
\begin{tabular}{l l l l l l l l l}
\toprule
& \multicolumn{2}{l}{SITW EVAL CORE} & \multicolumn{2}{l}{SITW EVAL MULTI} & \multicolumn{2}{l}{SRE18 EVAL VAST} & \multicolumn{2}{l}{SRE19 EVAL VAST} \\
\cmidrule(lr){2-3} \cmidrule(lr){4-5} \cmidrule(lr){6-7} \cmidrule(lr){8-9}
& EER & Min Cost & EER & Min Cost & EER & Min Cost & EER & Min Cost \\
\midrule
TDNN (mean \& std) & 2.21 & 0.135 & \textbf{3.46} & \textbf{0.183} & 12.69 & \textbf{0.472} & 5.97 & 0.223 \\
TDNN-SE (mean \& std) & 2.02 & 0.125 & 4.03 & 0.188 & 12.70 & 0.473 & 5.97 & 0.212 \\
TDNN-RES-SE (mean \& std) & 2.10 & 0.123 & 4.07 & 0.188 & 12.02 & 0.477 & 5.75 & 0.216 \\
TDNN-RES-SE (LDE, isotropic) & 2.02 & 0.122 & 4.04 & 0.185 & 12.70 & 0.497 & 5.53 & 0.212 \\
\rowcolor[RGB]{200, 212, 255} $\quad$ \nextline Neural i-vector & 2.93 & 0.173 & 5.55 & 0.249 & 15.92 & 0.588 & 6.64 & 0.254 \\
TDNN-RES-SE (LDE, shared diag.) & \textbf{1.83} & 0.123 & 4.42 & 0.189 & \textbf{11.75} & 0.483 & 5.34 & 0.213 \\
\rowcolor[RGB]{200, 212, 255} $\quad$ \nextline Neural i-vector & 2.81 & 0.168 & 5.40 & 0.246 & 15.87 & 0.522 & 6.43 & 0.256 \\
TDNN-RES-SE (NetVLAD) & 1.94 & \textbf{0.117} & 4.06 & 0.184 & 12.38 & 0.474 & \textbf{5.31} & \textbf{0.208} \\
\rowcolor[RGB]{200, 212, 255} $\quad$ \nextline Neural i-vector & 3.09 & 0.175 & 5.73 & 0.261 & 16.51 & 0.588 & 6.00 & 0.290 \\
TDNN-RES-SE (NetVLAD/LDE) & 2.02 & 0.129 & 4.41 & 0.199 & 13.40 & 0.528 & 5.53 & 0.229  \\
\rowcolor[RGB]{200, 212, 255} $\quad$ \nextline Neural i-vector & 3.06 & 0.188 & 5.65 & 0.262 & 15.56 & 0.596 & 5.97 & 0.253 \\
\bottomrule
\end{tabular}}
\end{table*}
\renewcommand*{\arraystretch}{1.0}
\addtolength{\tabcolsep}{1pt}  

\vspace{-1mm}
\subsection{Scoring back-end}
\vspace{-1mm}
We centered, whitened, and length-normalized (both discriminative and generative) speaker embeddings before simplified PLDA scoring \cite{garcia2011analysis}. We did not apply domain adaptation techniques, but simply used the training data (VoxCeleb) to compute centering vector and whitening matrix. Finally, we performed \emph{adaptive symmetric score normalization} (AS-norm)~\cite{cumani2011comparison}. For AS-norm, we randomly selected $2000$ utterances from training data and chose $200$ highest scoring utterances for each enrollment or test utterance to compute the normalization statistics.

\vspace{-1mm}
\subsection{Speaker verification results}
\vspace{-1mm}

\renewcommand*{\arraystretch}{1.1}
\begin{table}[th!]
\caption{{\it Review of recent single system results for SITW core-core condition. Due to different experimental settings and implementations, the results from different approaches are not directly comparable. Out of the i-vector systems, the proposed neural i-vector obtains the lowest EER. The second lowest EER was obtained by an i-vector system using a dereverberation system (WPE) together with perceptual linear prediction (PLP) and stacked bottleneck features (SBN). Other two included i-vector systems use MFCCs and bottleneck features (BNF). Under the divider line are the systems based on deep speaker embeddings. All systems use either MFCCs or filterbank coefficients (FBANK) as input features. All the embedding networks use either TDNN, extended TDNN (E-TDNN), factorized TDNN (F-TDNN), or ResNet34 based architectures. One system uses additive angular margin (AAM) loss instead of standard cross-entropy. The performance differences between the deep embedding extractors are rather small, except for the last system utilizing tied mixture of factor analyzers (TMFA) layer that is trained on 8 kHz Switchboard and SRE data.} \label{table:sitw_results}}
\vspace{-1mm}
\centering
\begin{tabular}{l l}
\toprule
System \& study & EER (\%) \\
\midrule
Neural i-vector [this study] & 2.81 \\
WPE PLP+SBN i-vector \cite{matejka2019} & 3.38 \\
MFCC i-vector \cite{matejka2019} & 4.40 \\
BNF i-vector \cite{villalba2020} & 5.77 \\
\midrule
TDNN-RES-SE (LDE) [this study] & 1.83 \\
FBANK E-TDNN \cite{matejka2019} & 1.70 \\
MFCC E-TDNN \cite{snyder2019speaker} & 1.7 \\ 
MFCC F-TDNN \cite{villalba2020} & 1.86 \\
FBANK ResNet34+LDE (AAM-softmax) \cite{villalba2020} & 2.11 \\
FBANK ResNet34+TMFA (8 kHz) \cite{chen2019tied} & 5.74 \\
\bottomrule
\centering
\vspace{-3mm}
\end{tabular}
\end{table}
\renewcommand*{\arraystretch}{1.0}

Table \ref{table:results} shows the results of our experiments with different systems on multiple speaker verification evaluations. The results for the core condition of SITW are the most representative of the basic accuracy of the ASV systems as it does not have multi-speaker utterances requiring diarization. The other evaluations provide supporting evidence, although the results may be impaired by the lack of diarization. 

In general, we find that the differences between the results of different deep embedding extractors are small. For example, when migrating from TDNN to TDNN-SE and to TDNN-RES-SE architectures, the results slightly improve on some evaluations, but get slightly worse on others.
Similarly, the differences between the different aggregation methods are relatively minor, which is quite intriguing considering the differences between the standard mean and standard deviation aggregation and the dictionary based methods.

Different variants of neural i-vectors perform almost equally well to each other. The performance of neural i-vectors is way behind the performance of their deep embedding counterparts. On the other hand, the neural i-vectors perform substantially better than the other i-vector systems reported in literature as can be observed from Table \ref{table:sitw_results}. The table also shows that our deep embeddings obtain a competitive results in comparison to the results reported in the other studies.

\subsection{Visualizations of neural i-vectors}
\vspace{-0.5mm}

\begin{figure}[b!]
  \centering
  \centerline{\includegraphics[width=0.96\linewidth]{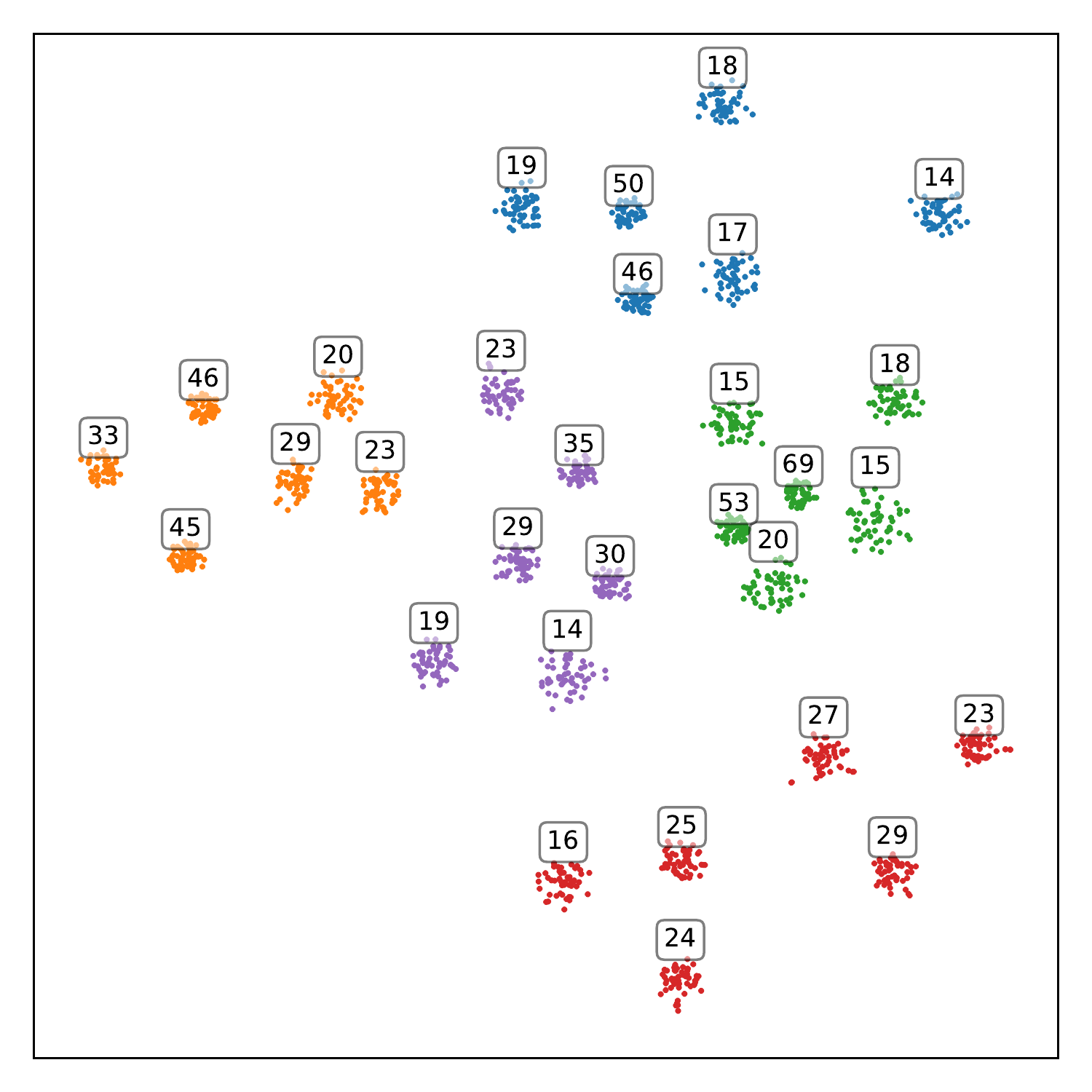}}
  \vspace{-3mm}
  \caption{{\it T-SNE visualization \cite{maaten2008visualizing} of random neural i-vectors drawn from i-vector posterior distributions of 30 utterances from 5 male speakers in SITW corpus. Different colors represent different speakers. Each of the 30 clusters consists of 50 random i-vectors drawn from the posterior distribution of one i-vector. The numbers show durations of the utterances in seconds after removing non-speech frames. The long utterances have less uncertainty than the short ones, which can be observed from the compactness of the clusters.}}
  \label{fig:sampled_ivectors}
\end{figure}

In Figure \ref{fig:sampled_ivectors}, we illustrate \emph{sampled} neural i-vectors for 5 male speakers in the SITW corpus. From each speaker we selected six utterances and computed the posterior distributions \cite[eqs. (3) and (4)]{vestman2019} of i-vectors. These distributions were used to sample 50 i-vectors per utterance. From the figure, we can observe that different speakers are well separated and that the utterances with short durations have higher uncertainty (\ie more spread clusters) than the utterances with long durations, as expected.

\begin{figure}[t!]
  \centering
  \centerline{\includegraphics[width=0.9\linewidth]{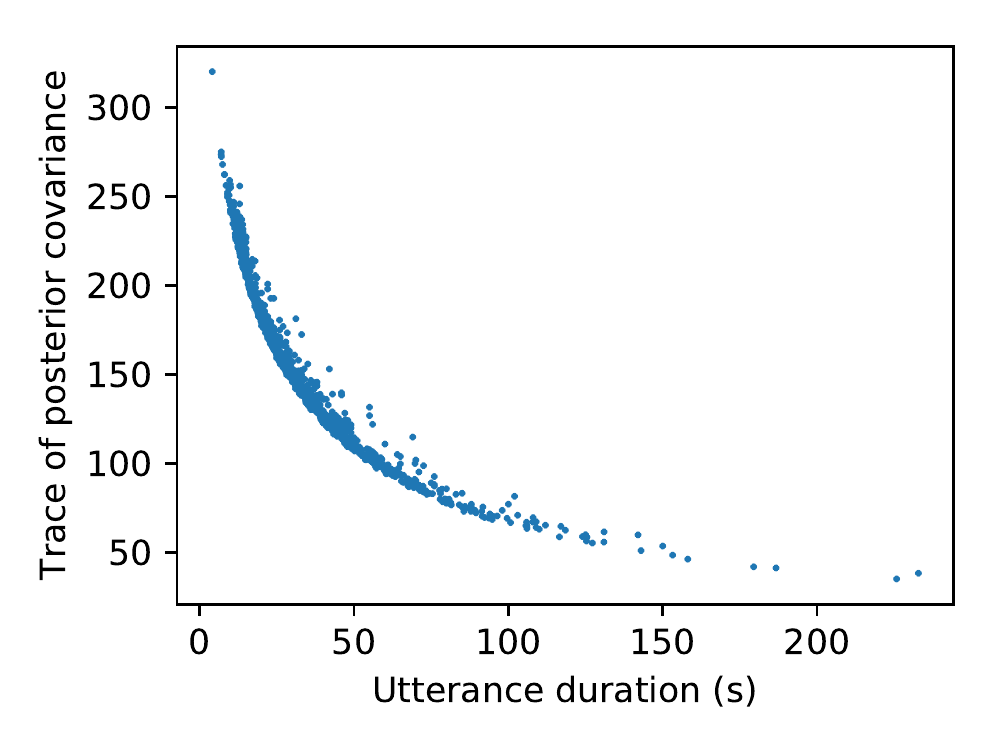}}
  \vspace{-3mm}
  \caption{{\it Trace of i-vector posterior covariance matrix as a function of utterance duration for utterances in SITW core-core condition.}}
  \label{fig:trace_vs_duration}
\end{figure}

Finally, in Figure \ref{fig:trace_vs_duration}, we depict traces of posterior covariances \cite[eq. (3)]{vestman2019} of i-vectors for SITW data. The traces reflect the uncertainty in the i-vector estimation \cite{poorjam2016incorporating}. As expected, the longer the duration, the less uncertain the i-vectors are.

\section{Conclusion}
\blfootnote{This work was partially supported by Academy of Finland (project 309629).}
At a broad outlook, the general developments in the field of speaker recognition have involved innovative (and often successful) re-use of previous generation tools to build up next generation recognizers: we have 
seen steady transition in state-of-the-art from individually trained GMMs to GMM-UBM, GMM supervectors, JFA and i-vectors (in this order). As a community, we have been working on multi-layered (deep) models, formed by stacking frame-level feature processors with utterance-level presentations and speaker latent variable models. Until the recent past, however, these pipelines have not been trained as a whole, but constructed from individually-optimized components. This is where the deep neural networks have come to a rescue, and we are witnessing transition towards the next generation deep models. Nonetheless, deep neural network models have seem to have interrupted the chain of GMM-based systems, particularly as they lack the concept of a universal background model. Some recent work has therefore looked into replacing the global temporal pooling operation of deep embedding extractors with learnable dictionaries, similar to the UBM, with demonstrated improvements.  

In an attempt to bridge classic GMM-based technology and the modern deep learning era, we have provided a unified comparison of alternative i-vector extractors that use different variants of deep neural networks to optimize the frame-level features and the UBM. In particular, two recent deep neural network architectures, LDE and NetVLAD, \emph{can be interpreted as GMMs with specific assumptions}. This interpretation enabled us to re-consider classic GMM-based systems using discriminatively obtained features and UBM. As a proof of concept, we decided to focus on the i-vector system, but similar construction is readily applicable to any ASV or diarization system that uses GMMs.

Our results indicate that `neural i-vectors' outperform all the existing i-vector variants by a wide margin, indicating the importance of using speaker-informative short-term features and speaker-informative dictionary. Even if the corresponding `purely neural' systems (used for obtaining the components of our i-vector system) outperform the neural i-vector approach, this was \emph{not} the point of our study. The point, instead, is that it is possible to view certain neural architectures as if having a multi-modal aggregator (GMM) built in them. These identified connections may open up fresh ideas in revoking techniques such as uncertainty propagation, data augmentation (by sampling features or speaker embeddings). Potential applications that may benefit from fine-grained frame-by-frame speaker decisions, such as speaker diarization, provide another potential topic of future studies.

\bibliographystyle{IEEEbib}
\bibliography{references}

\end{document}